\documentclass[aps,prd,nofootinbib,twocolumn,superscriptaddress,showpacs,showkeys]{revtex4-1}
\usepackage {amsmath}
\usepackage {amssymb}
\usepackage {amsfonts}
\usepackage {amsthm}
\usepackage {mathrsfs}
\usepackage {natbib}
\usepackage {latexsym}
\usepackage {graphicx}
\usepackage {dsfont}
\usepackage {times}
\usepackage {rotating}
\usepackage {wasysym}
\usepackage {multirow}
\usepackage {hhline}
\usepackage {hyperref}
\usepackage {color}
\usepackage {bm}
\usepackage{appendix}
\usepackage{acronym}
\usepackage{dcolumn}   
\usepackage {url}
\usepackage[normalem]{ulem}   

\hypersetup{
  colorlinks=true,        
  linkcolor=blue,         
  citecolor=cyan,         
}

\def\commitID{commitID: 01ffd337f9a75416522dfe67ead791f8117ce229}
\def\commitDATE{Thu Dec 5 17:14:10 2013 +0100}

\newcommand{\dcc}{LIGO-P1300211-v1}

\newcommand{\Msun}{M_{\odot}}

\newcommand{\cf}{{cf.}~}

\begin{document}

\title{Host redshifts from gravitational-wave observations of binary neutron star mergers}
\author{C.~Messenger}
\affiliation{School of Physics and Astronomy,University of Glasgow, University Avenue, Glasgow, G12 8QQ, UK}
\author{Kentaro~Takami}
\affiliation{Max-Planck-Institut f{\"u}r Gravitationsphysik, Albert Einstein
Institut, Am M{\"u}hlenberg 1, 14476 Potsdam, Germany}
\affiliation{Institut f{\"u}r Theoretische Physik, Max-von-Laue-Str. 1, 60438
Frankfurt, Germany}
\author{Sarah~Gossan}
\affiliation{TAPIR, California Institute of Technology, 1200 E
  California Blvd., Pasadena, CA 91125, USA}
\author{Luciano~Rezzolla}
\affiliation{Institut f{\"u}r Theoretische Physik, Max-von-Laue-Str. 1, 60438
Frankfurt, Germany}
\affiliation{Max-Planck-Institut f{\"u}r Gravitationsphysik, Albert Einstein
Institut, Am M{\"u}hlenberg 1, 14476 Potsdam, Germany}
\author{B.~S.~Sathyaprakash}
\affiliation{School of Physics and Astronomy, Cardiff University, 5, The
  Parade, Cardiff, UK, CF24 3AA}
%

\date{\today}
\date{\commitDATE\\\mbox{\small \commitID}\\\mbox{\dcc}}

\begin{abstract}
  Inspiralling compact binaries as standard sirens will soon become an
  invaluable tool for cosmology when advanced interferometric
  gravitational-wave detectors begin their observations in the coming
  years. However, a degeneracy in the information carried by
  gravitational waves between the total rest-frame mass $M$ and the
  redshift $z$ of the source implies that neither can be directly
  extracted from the signal, but only the combination $M(1+z)$, the
  redshifted mass. Recent work has shown that for binary neutron star
  systems, a tidal correction to the gravitational-wave phase in the
  late-inspiral signal that depends on the rest-frame source mass could
  be used to break the mass-redshift degeneracy. We propose here to use
  the signature encoded in the post-merger signal to deduce the redshift
  to the source. This will allow an accurate extraction of the intrinsic
  rest-frame mass of the source, in turn permitting the determination of
  source redshift and luminosity distance solely from gravitational-wave
  observations. This will herald a new era in precision cosmography and
  astrophysics. Using numerical simulations of binary neutron star
  mergers of very slightly different mass, we model gravitational-wave
  signals at different redshifts and use Bayesian parameter estimation to
  determine the accuracy with which the redshift can be extracted for a
  source of known mass. We find that the Einstein Telescope can determine
  the source redshift to $\sim 10$--$20\%$ at redshifts of $z<0.04$.
\end{abstract}

\maketitle

\acrodef{GW}[GW]{gravitational-wave}
\acrodef{BNS}[BNS]{binary neutron star}
\acrodef{HMNS}[HMNS]{hypermassive neutron star}
\acrodef{ET}[ET]{Einstein Telescope}
\acrodef{NS}[NS]{neutron star}
\acrodef{EM}[EM]{electromagnetic}
\acrodef{SNR}[SNR]{signal-to-noise ratio}
\acrodef{PDF}[PDF]{probability distribution function}
\acrodef{EOS}[EOS]{equation of state}

\section{Introduction}
%
The prospects for \ac{GW} astronomy in the era of
advanced detectors are promising, with several detections expected before
the end of the decade when Advanced LIGO~\citep{Harry:2010}, Advanced
Virgo~\citep{Virgo:2009} and KAGRA~\citep{Aso:2013} become fully
operational. Among the sources of \acp{GW} expected to be detected are the
inspiral and coalescence of \acp{BNS}, neutron
star-black hole binaries, and binary black holes.  Population models
suggest that the detection rate of compact binary coalescences for \acp{BNS}
will be $\sim 10\,\mathrm{yr}^{-1},$ when Advanced LIGO
\cite{Abadie:2010cf} reaches its design sensitivity.

The inspiral of compact binary systems are also known as \emph{standard
  sirens}~\citep{Schutz:1986}, as their luminosity distance can be
extracted from \ac{GW} observations alone, without the need for any detailed
modelling of the source, or of the properties of the media along the GW
path. This is because the observed amplitude of \acp{GW} during the inspiral
phase reaches the detector essentially unaltered and depends on a small
number of parameters, which can all be measured using a network of GW
detectors. These parameters include the total gravitational mass and mass
ratio of the system, the spins of the compact objects, the orientation of
the binary's orbital plane with respect to the line of sight, the
source's position on the sky and the luminosity distance to the
source. \ac{GW} observations can very accurately measure the signal's phase
evolution, which depends only on the total mass and mass ratio of a
binary. Simultaneously, a network of detectors can determine the sky
position, polarisation amplitudes and the distance to the binary. The
observed total mass, however, is not the system's intrinsic mass $M$
(i.e., mass as measured in the rest frame of the source) but the
\emph{redshifted} mass $M_z \equiv M(1+z)$. This is known as the
mass-redshift \emph{degeneracy}.

The mass-redshift degeneracy is detrimental to the application of GW
observations for cosmological inference. The relationship of the source's
luminosity distance to its redshift on cosmological scales is precisely
that which allows us to probe the parameters governing a cosmological
model. Breaking the mass-redshift degeneracy requires an electromagnetic
identification to tie the source to its host galaxy and thereby extract
the source's redshift. It was thought, until recently, that there is no
way to infer the source's redshift from \ac{GW} observations alone.

To use of \ac{GW} observations to extract information that is necessary for
cosmography [e.g., estimation of the Hubble parameter and the dark
energy \ac{EOS}] and astrophysics [e.g., measurement of the
  masses and radii of \acp{NS} and the \ac{EOS} of matter at
  supranuclear densities], requires precision measurements of both the
luminosity distance and intrinsic mass of the source. The mass-redshift
degeneracy forces reliance on electromagnetic identification of host
galaxies~\citep{2013arXiv1307.2638N,HolzHughes05,Sathyaprakash:2009xt,
  Nissanke:2009kt,Zhao:2011,Chen:2012}, which may be possible only very
rarely. For example, using gamma-ray bursts for identification of the
host galaxy greatly reduces the available signal population for
cosmography and could potentially lead to observational bias. This is
because gamma-ray emission is believed to be strongly beamed along a
jet~\cite{Nakar:2007yr,Lee:2007js,Rezzolla:2011}, while \ac{GW} emission is
expected to be approximately isotropic (quadrupolar) and hence only a
small fraction ($\sim 10^{-3}$) of all \ac{GW} events will have gamma-ray
burst counterparts \citep{Sathyaprakash:2009xt}. Determining the
electromagnetic counterparts to binary-black hole mergers is also a very
active area of research and several simulations have already been
performed in this context~\cite{Bode:2009mt,Farris:2009mt,Zanotti2010,
  Palenzuela:2010, Moesta2011} to provide first estimates on the
properties and energetics of these emissions.

Some authors have explored other approaches to measure cosmological
parameters without the aid of electromagnetic counterparts. These methods
either assume that the intrinsic \ac{NS} mass is constrained to be in a small
range around $1.4\,\Msun$~\citep{1993ApJ...411L...5C,
  1996PhRvD..53.2878F, Taylor:2013, Taylor:2012} or make use of a
statistical approach to measure the cosmological parameters from a
population of events~\citep{DelPozzo:2011yh}, as originally proposed by
Schutz~\citep{Schutz:1986}.
For \ac{BNS} systems, however, there are two signatures in the GW signal
from the late-inspiral and post-merger stages that depend on the
rest-frame source mass and could potentially provide a measure of the
source's intrinsic mass. Such a measurement would break the
mass-redshift degeneracy and help return both the source redshift and
intrinsic total mass from \ac{GW} observations alone. The first effect
is concerned with the correction in the orbital phase due to tidal
effects that appear at order $(v/c)^{10}$ beyond the leading order in
the post-Newtonian approximation to Einstein's
equations~\citep{Flanagan:2007}, where $v$ and $c$ are the orbital and
light velocities, respectively. This is a secular effect that becomes
important as the two bodies approach each other and was first proposed
as a cosmological tool by~\cite{Messenger:2011gi}. The second effect
occurs in the post-merger stage and causes a significant departure in
the post-Newtonian evolution of the system. Unless the two stars are
not very massive, the newly formed object is a \ac{HMNS}
\cite{Baiotti:2008}, which develops a bar-mode deformation, which can
survive even for a fraction of a second (\cf Fig.~A1
of~\cite{Rezzolla:2010}), delaying the birth of the black hole and
emitting \acp{GW} in a narrow frequency range. The importance of this
effect has not yet been considered for cosmological exploitation.

\begin{figure}
\centering
\includegraphics[width=\columnwidth]{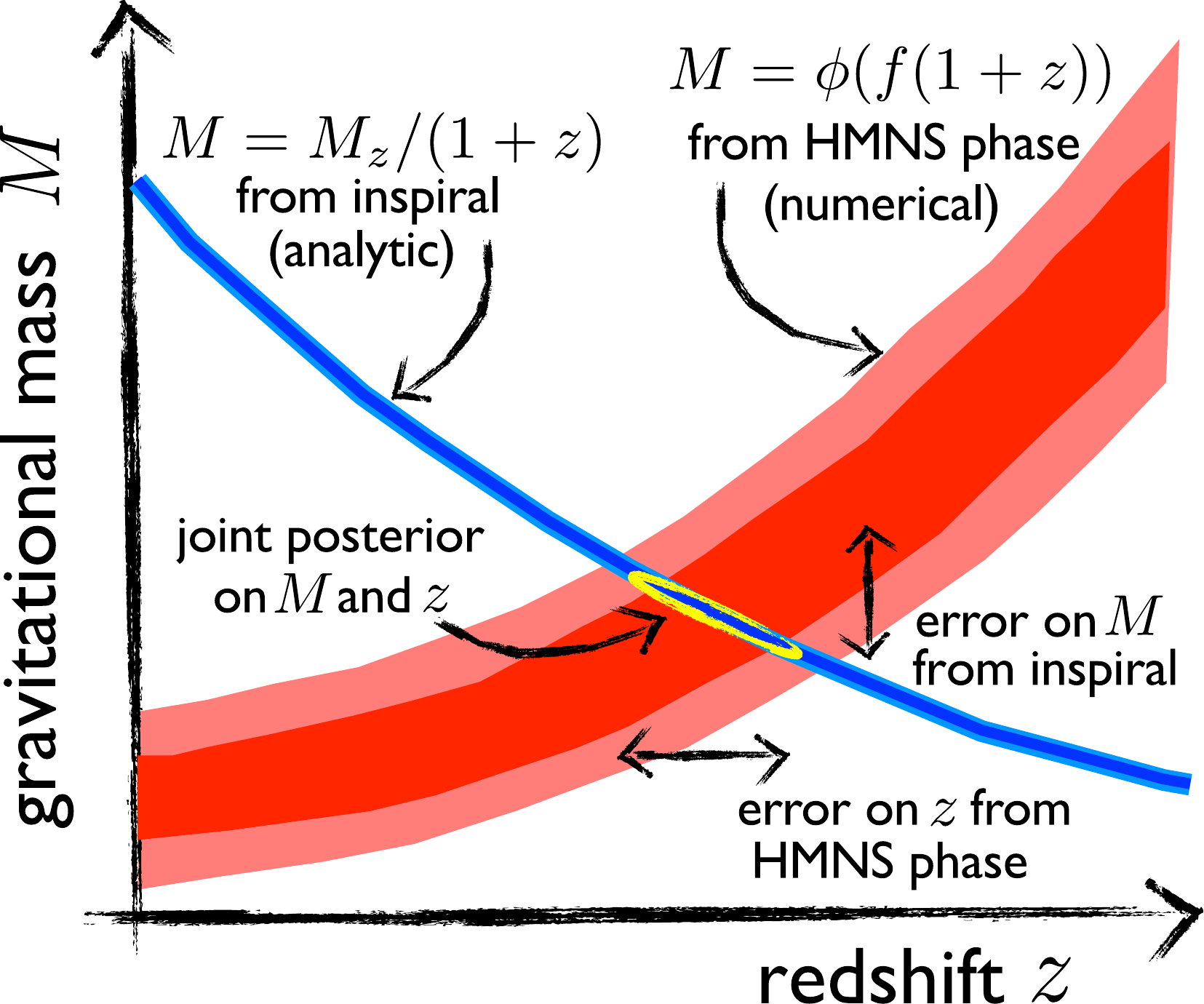}
\caption{A cartoon of how the mass-redshift degeneracy is broken
  through the use of information from the inspiral and \ac{HMNS} stage
  of a \ac{BNS} merger event. Information on the redshifted mass as a
  function of the the redshift (blue stripe) can be correlated with
  complementary information from the spectral properties of the
  \ac{HMNS} phase. The overlap will provide a localised range in mass
  and redshift, breaking the degeneracy.\label{fig:sketch}}
\end{figure}

In this paper, we propose to use the \ac{GW} signal including the
inspiral phase and the \ac{HMNS} signature to extract the intrinsic
gravitational masses and the source redshift. The power spectrum of the
\ac{HMNS} stage of the merger of \ac{BNS} systems has been shown to
contain prominent spectral features that vary smoothly in frequency with
the total gravitational mass of the
system~\citep{Bauswein2011,Takami2014}. The inspiral stage of the
waveform can be used to obtain a highly accurate measurement of the
\emph{redshifted} total mass of the system $M_{z}$. This allows us to
constrain the true values of the rest mass and redshift to a relatively
narrow band spanning the full range of the $(z,M)$ plane.  Independently,
the \ac{HMNS} stage of the waveform allows us to measure the
\emph{redshifted} fundamental frequencies of two prominent spectral
features to a reasonable accuracy. Using these and an empirically
determined relationship between the total gravitational mass of the
binary and the rest-frame fundamental frequencies, we are then
able to independently constrain a second region of the $(z,M)$ plane.
The localised intersection of the two regions in the $(z,M)$ plane allows
us to break the mass-redshift degeneracy present in both measurements and
make an estimate of the redshift \emph{and} gravitational mass of the
system. A cartoon of this idea is shown in Fig.~\ref{fig:sketch}, which
can be directly compared to an example of the results of this analysis
shown in Fig.~\ref{fig:zM_multimap}.
%

The detectability of the \ac{HMNS} part of the waveform is not likely to be
high for sources observed in advanced detectors, as these features lie at
frequencies significantly higher than the sensitive bands of ground based
detectors. We will, therefore, explore how we might use the signature of
\ac{HMNS} in the context of the \ac{ET}, a third-generation
ground-based interferometric \ac{GW} detector~\citep{Punturo:2010}.

The rest of the paper is organised as follows: In
Section~\ref{sec:numerical} we describe the numerical waveforms used for
this analysis. In Section~\ref{sec:modelling} we describe our robust, but
ad-hoc, parameterisation and modelling of the \ac{HMNS} power spectrum. In
Section~\ref{sec:analysis} we describe the analysis methods used to
simulate and measure the \ac{HMNS} spectral features. We then describe the
procedure with which these measurements are combined to obtain the
redshift and gravitational masses of the source. Finally, in
Section~\ref{sec:conclusions} we conclude with discussions of our results
and future directions for this research.

\section{Numerical simulations of BNSs}
\label{sec:numerical}

All of our calculations have been performed in full general relativity.
The evolution of the spacetime is obtained by using the \texttt{CCATIE}
code, a finite-differencing code providing the solution of a conformal
traceless formulation of the Einstein equations~\citep{Pollney:2007ss},
with a ``$1+\log$'' slicing condition and a ``Gamma-driver'' shift
condition. The general-relativistic hydrodynamics equations are solved
using the \texttt{Whisky} code~\citep{Baiotti04,Baiotti:2008}, with the
Marquina flux formula and a PPM reconstruction. For the sake of
simplicity we model the \ac{NS} matter as an ideal fluid with a gamma-law \ac{EOS},
$p=(\Gamma -1)\rho\epsilon$ with $\Gamma=2$, where $p$ is the pressure,
$\rho$ the rest-mass density, and $\epsilon$ specific internal energy
(see ~\cite{Rezzolla_book:2013} for details).
The grid hierarchy, with a reflection symmetry condition across the $z=0$
plane and a $\pi$-symmetry condition across the $x=0$ plane, is handled
by the \texttt{Carpet} mesh refinement driver~\citep{Schnetter-etal-03b},
where we use 6 refinement level and the spacing of the finest grid is
$0.15\,G\Msun c^{-2} \sim 0.221\,\mathrm{km}$. We extract the \acp{GW},
consisting of a plus and cross polarisation and sampled in time at a rate
of $\Delta t=1.68\,G\Msun c^{-3} \sim 8.27\times10^{-3}\,\mathrm{ms}$
equivalent to a sampling rate of $\sim 121\,\mathrm{kHz}$, at a distance
$R_{0}=500\,G\Msun c^{-2} \sim 738\,\mathrm{km}$. We analyse only the
$\ell=m=2$ mode of \acp{GW}, which is the dominant one.
As initial data, we use quasi-equilibrium irrotational \acp{BNS} generated by
the multi-domain spectral-method code
\texttt{LORENE}~\citep{Gourgoulhon-etal-2000:2ns-initial-data} under the
assumption of a conformally flat spacetime metric. We have considered
five equal-mass binaries with an initial coordinate separation of the
stellar centres of $45\,\mathrm{km}$, and polytropic \ac{EOS}, $p = K \rho
^{\Gamma}$ with an adiabatic exponent $\Gamma=2$ and polytropic constant
$K=123.6$ (in units $c = G = \Msun = 1$); details on the different
binaries are collected in Table \ref{tab:ID}. A very important
requirement of our sample of \acp{BNS} is that they are only very finely
separated in total gravitational mass, with differences that are of the
order of $2\%$ only. Producing such a sample at a fixed separation is far
from trivial and has represented a major numerical difficulty, stretching
the capabilities of the \texttt{LORENE} libraries. Once evolved, the
stars perform approximately 3.5 orbits before merger.

\begin{table}
\begin{tabular}{cccccc}
\hline \hline
$M_\mathrm{b}$ & $M_\mathrm{ADM}$ 
& $M_\infty$ & $R_\infty$ 
& $\mathcal{C}$ & $f_\mathrm{orb}$ \\
\hline
$[\Msun]$ & $[\Msun]$ 
& $[\Msun]$ & $[G\Msun/c^{2}]$ 
& & $[\mathrm{Hz}]$ \\
\hline
1.4237 & 2.6578 & 1.3413 & 11.386 & 0.11781 & 281.80 \\
1.4662 & 2.7305 & 1.3784 & 11.276 & 0.12224 & 284.62 \\
1.5099 & 2.8049 & 1.4163 & 11.158 & 0.12693 & 287.45 \\
1.5549 & 2.8811 & 1.4550 & 11.031 & 0.13190 & 290.29 \\
1.5947 & 2.9478 & 1.4890 & 10.914 & 0.13643 & 292.74 \\
\hline \hline
\end{tabular}
\caption{Properties of our initial data of equal-mass \acp{BNS} with the
  initial coordinate separation $45\,\mathrm{km}$. Reported in the
  various columns are the baryon mass $M_\mathrm{b}$ of each star, the
  ADM mass $M_\mathrm{ADM}$ of the system at initial, the gravitational
  mass $M_\infty$ of each star at infinite separation ($M=2M_\infty$),
  the circumferential radius $R_\infty$ of each star at infinite
  separation, the compactness ${\mathcal C}\equiv M_\infty/R_\infty$ and
  the orbital frequency $f_\mathrm{orb}$ at the initial separation.
\label{tab:ID}}
\end{table}

\section{Frequency domain modelling of the HMNS}
\label{sec:modelling}

In order to perform the parameter estimation described in
Sec.~\ref{sec:analysis} we must first be able to parametrise and model
the \ac{HMNS} stage of the waveform. Using our five waveforms as a basis, the
current state-of-the-art numerical simulations of \ac{BNS} systems do not yet
give us the insight and accuracy required to model the phase evolution of
the \ac{HMNS} waveform as a function of the system's mass. This, coupled with the
assumption that there exists a smooth relationship between the total
gravitational mass of the system $M$ and the frequencies of prominent
spectral features, forces us to model the signal \emph{power} rather than
the complex waveform. Therefore, in our \ac{HMNS} analysis, we are insensitive
to information encoded in the phasing of the waveform. Unless a
semi-analytic description of the phase evolution in the \ac{HMNS} stage is
possible, the one adopted here is probably the only approach feasible.

For each numerical waveform we perform the following procedure in order
to compute noise-free power-spectrum reference templates. The time series
for both the plus and cross polarisations are pre-processed using a
fifth-order high-pass Butterworth filter with knee-frequency $1$ kHz and
a time-domain Tukey window with alpha parameter $0.25$. This is done to
suppress the leakage of power from the last few cycles of the inspiral
and initial merger stage of the waveform. The discrete Fourier transform
is then computed for each polarisation from which we construct the
reference template
\begin{equation}\label{eq:reftemplate}
\mathcal{T}(f) \equiv \frac{|\tilde{h}_{+}(f)|^{2}+
|\tilde{h}_{\times}(f)|^{2}}{S_{h}(f)}\,,
\end{equation}
where $S_{h}(f)$ is the noise spectral density of the detector, which we
choose to be that of the ET-B~\cite{Hild:2008} design\footnote{Using the
  more recent ET-D design enhances the sensitivity in the region of
  interest $1$--$3$ kHz by $\approx 12\%$.}.

\begin{figure}
\centering
\includegraphics[width=\columnwidth]{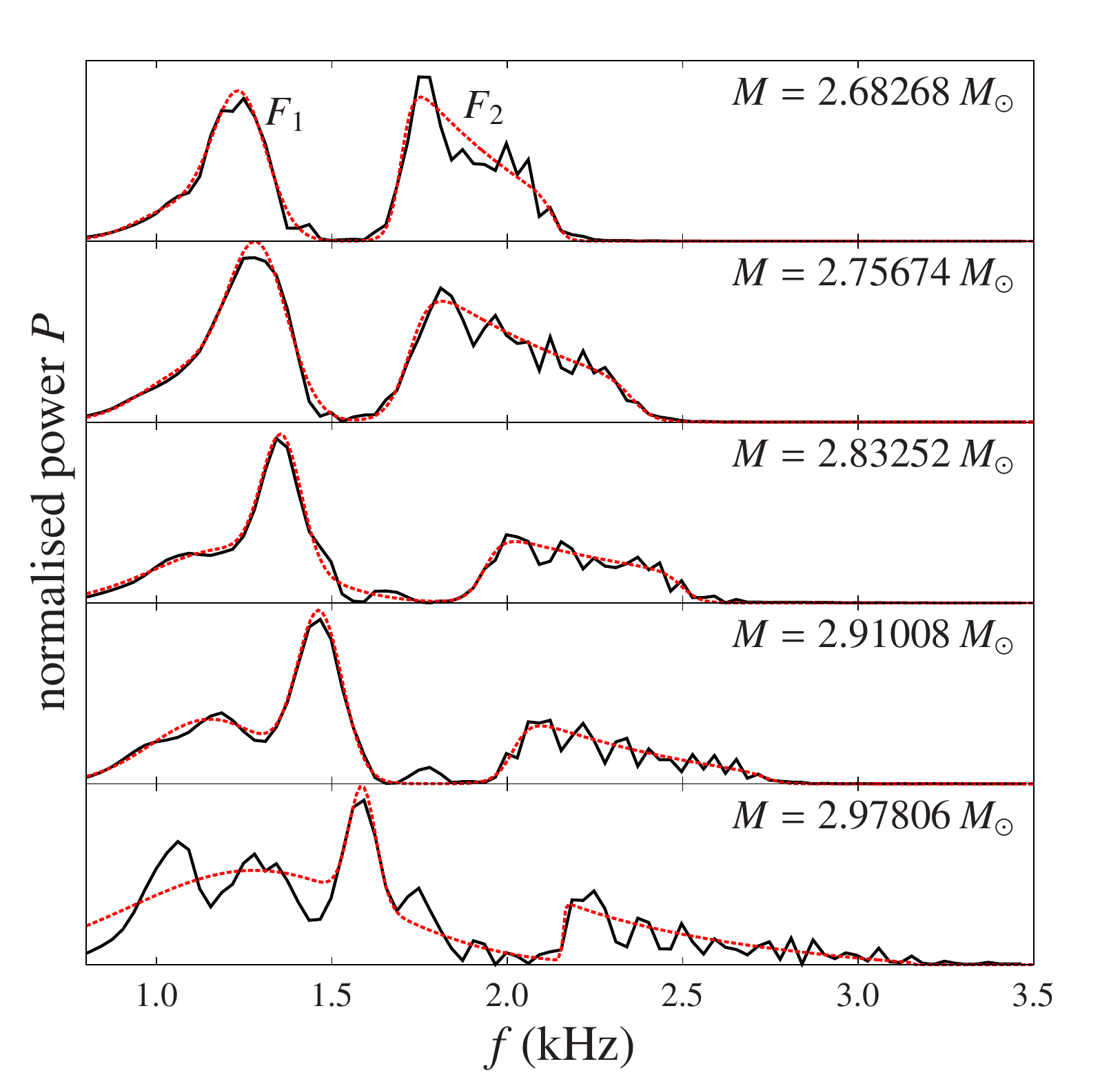}
\caption{The normalised power-spectrum reference templates
  [Eq.~\eqref{eq:reftemplate}] for each of the five system masses as a
  function of frequency (black lines). Also plotted are the best fit
  model templates defined in Eq.~\eqref{eq:template} (red dashed lines).}
\label{fig:power}
\end{figure}

Visual inspection of these reference templates as a function of
frequency, shown in Fig.~\ref{fig:power}, allows us to clearly identify
the two primary spectral features of interest. The first feature, at
frequencies $\approx1.2$--$1.6$ kHz is approximately Gaussian in profile
and moves to higher frequencies for higher mass systems. In contrast, the
second feature, at frequencies $\approx1.7$--$3$ kHz, appears to be best
described by a sloping trapezoid with rounded shoulders and a central
frequency and bandwidth that also grows with increasing system mass. In
addition, there appears to be a third power component at low frequencies,
i.e., $\leq 2$ kHz, that becomes more dominant as the system mass
increases. A reasonable approximation to this third feature is a second
Gaussian of lower amplitude and greater variance than that used to model
the first feature. For the purposes of this work, this third feature is
included only to improve the quality of our model
fitting. Mathematically, our entire ad-hoc model of the waveform
power-spectrum can be expressed as
\begin{align}\label{eq:template}
  \mathcal{S}(f;\bm{\lambda}) &= S_{h}^{-1}(f)\biggl(A_{1}e^{-(f-F_{1})^2/W_{1}^{2}}+ A_{3}e^{-(f-F_{3})^2/W_{3}^{2}} \nonumber\\ 
  &+A(f;A_{2a},A_{2b},F_{2},W_{2})\gamma(f;F_2,W_2,s)\biggr)\,,
\end{align}
where $A_{1}$ and $A_{3}$ are amplitude terms with $F_{1},F_{3}$ and
$W_{1},W_{3}$ as central frequencies and half-widths (standard
deviations), respectively, of the Gaussian features. The template is
whitened using the detector noise spectral density as done for the
reference template. We also define 
\begin{align}
\label{eq:slope}
  & A(f;A_{2a},A_{2b},F_{2},W_{2}) \equiv \nonumber \\
& \ \ \frac{1}{2W_{2}}
\left[
\left(A_{2b}-A_{2a}\right)\left(f-F_{2}\right)+W_2\left(A_{2b}+A_{2a}\right)
\right]\,,
\end{align}
which is a linear slope of amplitude $A_{2a}$ for $f=F_{2}-W_{2}$ and
amplitude $A_{2b}$ for $f= F_{2}+W_{2}$. Finally, the function
\begin{align}
\label{eq:shoulder_0}
& \hskip -0.75cm \gamma(x;F_{2},W_{2},s) \equiv \nonumber \\
& \frac{1}{1+e^{-(f-F_{2}+W_{2})/s}} - 
\frac{1}{1+e^{-(f-F_{2}-W_{2})/s}}\,,
\end{align}
is the difference between two simple sigmoid functions which serves to
bound our model of the second spectral feature component between the
frequencies $F_{2}\pm W_{2}$ with a smooth transition from zero to unity
over a fixed frequency range controlled by the parameter $s$. This ad-hoc
template is therefore described by the 11-dimensional parameter vector
$\bm{\lambda} \equiv
(A_{1},A_{2a},A_{2b},A_{3},F_{1},F_{2},F_{3},W_{1},W_{2},W_{3},s)$.

We employ a simple least-squares fitting procedure to obtain our best fit
parameters $\bm{\lambda}'(M)$ for each system mass. In all cases there
were restrictions on the allowed parameter space ensuring that: (a)
$A_{2a}>A_{2b}$, such that the slope of the second spectral feature was
negative; (b) $F_{3}<F_{2}$, such that the third (Gaussian) spectral
feature was restricted to the lower frequencies; (c) the smoothing length
$s<W_{2}/5$, such that the smoothed transition regions of the second
feature account for less that $10\%$ of the total feature width. As can
be seen from Fig.~\ref{fig:power}, our choice of model and parameter
restrictions provides a reasonable fit to the numerical data across our
range of masses. We note that the quality of fit does begin to
deteriorate at higher masses. This is due to the fact that as the mass of
the binary increases, the \ac{HMNS} is further away from a stable
equilibrium and its dynamics is much more violent; in particular, the
bar-deformed object rapidly spins up via the copious emission of
\acp{GW}, leading to a very broad spectrum for the $F_2$ frequency.

\begin{figure}
\centering
\includegraphics[width=\columnwidth]{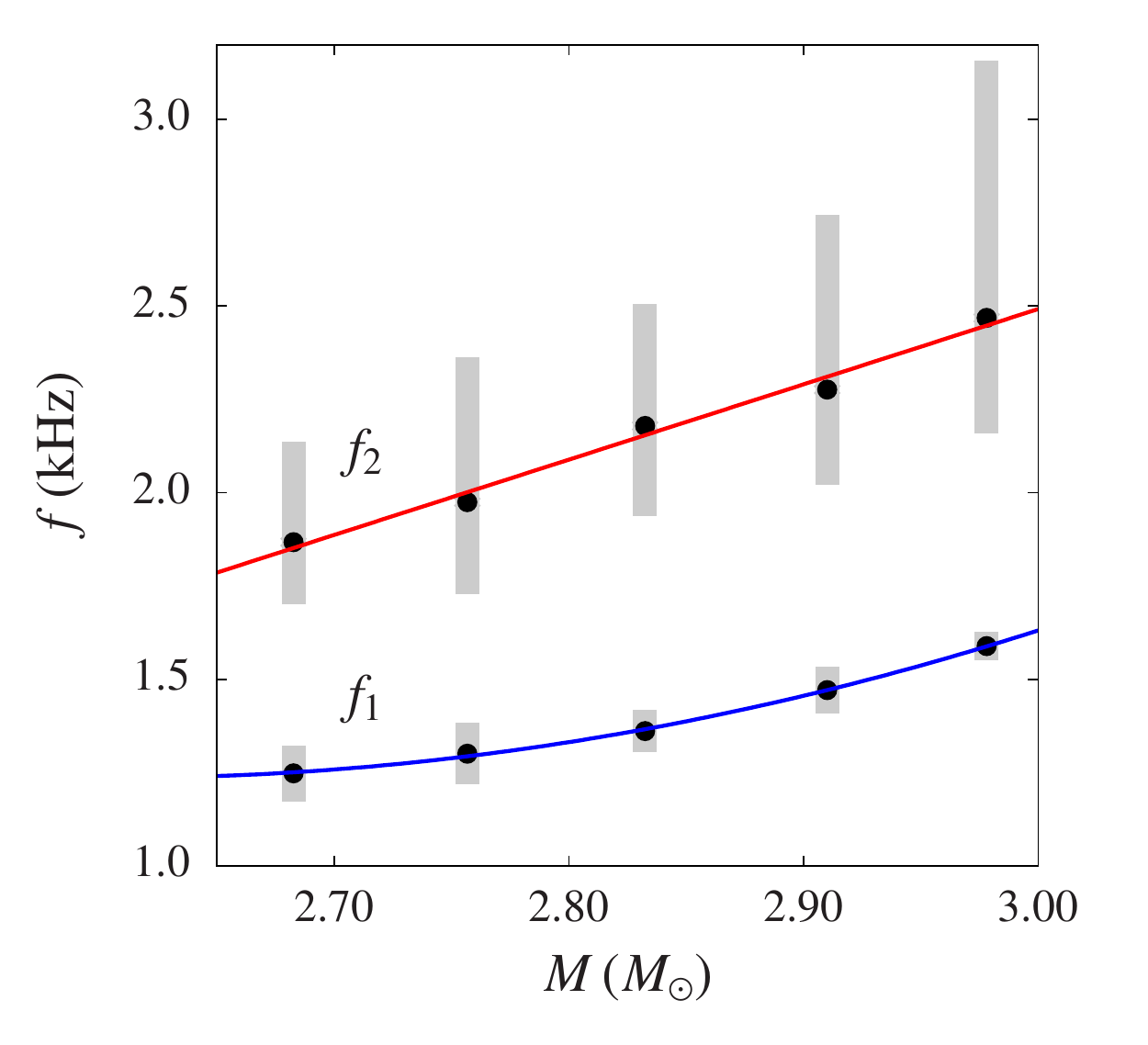}
\caption{Measured best fit values of $f_{1}$ and $f_{2}$ versus the
  total system mass (solid black circles). The vertical grey
  bars on the lower frequency spectral feature represent the standard
  deviation of the Gaussian fit at each mass value (not the
  measurement uncertainty). The corresponding bars on the higher
  frequency data represents the total frequency span of the feature. The blue
  and red curves show the least-squares fit to $f_{1}$ and $f_{2}$
  for a second and first-order polynomial respectively [given in
  Eq.~\eqref{eq:polyfit}].}
\label{fig:Mf}
\end{figure}

The model fit depends on the parameter set $\bm{\lambda}$, but we are
only truly interested in a measure of the characteristic frequencies
corresponding to the two dominant spectral features. For the
lower-frequency Gaussian feature we choose to use the central frequency
of the corresponding fit, $F_{1}$, as this measure. For the
higher-frequency, less symmetric, second feature we choose to define its
characteristic frequency as the average frequency within the bandwidth of
the second feature weighted by our best fit power model. This choice was
made in an attempt to more robustly track the location of the power of
the second feature. Hence, the lower and upper characteristic frequencies
are defined as
\begin{subequations}
\label{eq:f1f2}
\begin{align}
  f_{1} &\equiv F_{1}\,,\\
  f_{2} &\equiv 
 \frac{\int\limits_{0}^{\infty} 
 A(f;A'_{2a},A'_{2b},F'_{2},W'_{2})\,\gamma(f;F'_2,W'_2,s') f\,df}
 {\int\limits_{0}^{\infty} 
 A(f;A'_{2a},A'_{2b},F'_{2},W'_{2})\,\gamma(f;F'_2,W'_2,s') \,df}\,.
\end{align}
\end{subequations}
Taking the best fit parameters for each system mass and computing the
corresponding $f_{1}$ and $f_{2}$ values, we indeed validate the initial
hypothesis that there is a smooth relationship between these values and
the total gravitational mass of the system. This relationship is shown in
Fig.~\ref{fig:Mf} where we plot total gravitational mass versus $f_{1}$
and $f_{2}$. Also plotted are the following best-fit second and
first-order polynomials for $f_{1}$ and $f_{2}$, respectively, and whose
expressions in Hz are 
\begin{subequations}
\label{eq:polyfit}
\begin{align}
  f_{1}(M) &= 1331 + 992\,\Delta M + 2538\,\Delta M^2\,,\\
  f_{2}(M) &= 2087 + 2018\,\Delta M,
\end{align}
\end{subequations}
where $\Delta M/\Msun \equiv M/\Msun -2.8$. In all cases the residuals
from this polynomial fit are $<50$ Hz.  These functions represent our
empirically determined relationship between the characteristic
frequencies of the two spectral features and the total gravitational
mass. These phenomenological frequencies essentially track the
eigenfrequencies of the \ac{HMNS}, which has been recently shown to
behave as an isolated, self-gravitating system, whose dynamics can be
described as a superposition of different oscillation modes (see
\cite{Stergioulas2011b} for details).  

Some remarks are worth making. First, expression~\eqref{eq:polyfit} are
clearly tuned to our choice of \ac{EOS}, but equivalent expressions can be
derived for any \ac{EOS}~\cite{Takami2014}. Second, the functional dependence
of the frequencies on the total mass of the system is only weakly
dependent on the mass ratio; as a result, although our simulations refer
to equal-mass systems, we expect the functions~\eqref{eq:polyfit} to be a
good approximation also for unequal-mass binaries (see also
\citep{Bauswein2011} where this is discussed in detail). Finally, as
anticipated in the Introduction, the importance of the
relations~\eqref{eq:polyfit} is that they can be employed to break the
mass-redshift degeneracy.


\section{Analysis of the data}\label{sec:analysis}

We now describe how we simulate the occurrence and subsequent measurement
of \ac{BNS} signals in third-generation \ac{GW} detectors. We separate
this process into two parts. The first is the simulation and measurement
of the redshifted characteristic frequencies of the \ac{HMNS} spectral
features in the presence of Gaussian detector noise. The second is the
independent measurement of redshifted mass parameters using the inspiral
stage of a \ac{BNS} waveform. We then show how these measurements can be
combined to infer the gravitational mass and redshift of a source using
the results of the previous section.

\subsection{Measurement of the HMNS phase}\label{sec:barmode}

The numerical waveforms comprise time series of the plus and cross
polarisations of the \ac{GW} signal at a distance of $R_{0}=500\,\Msun$
which we label $h^{(0)}_{+}(t)$ and $h^{(0)}_{\times}(t)$
respectively. Using this data, we are able to simulate waveforms from
\ac{BNS} systems with arbitrary orientations, sky position, GW
polarisation, phase and redshift.

We first compute the redshifted and distance-scaled polarisations
\begin{subequations}
  \begin{align}
    \tilde{h}_{+}(f,z)& \equiv
\frac{R_{0}}{D_{_L}(z)}\tilde{h}^{(0)}_{+}\left(f(1+z)\right) \,,\\
    \tilde{h}_{\times}(f,z)& \equiv
\frac{R_{0}}{D_{_L}(z)}\tilde{h}^{(0)}_{\times}\left(f(1+z)\right)\,,
  \end{align}
\end{subequations}
where $D_{_L}$ is the luminosity distance and is a function of the
redshift $z$ and a choice of cosmological parameters. We use a standard
cosmological model described by the parameters: $h_{0}=0.71,$ $\Omega_{m} =
0.27,$ and $\Omega_{\Lambda} = 0.73$. In order to evaluate these waveforms at
arbitrary redshifted frequencies we use a cubic spline interpolation
scheme on the over-resolved discrete Fourier transform of the original
time series data.

To simulate \acp{GW} with arbitrary sky positions and nuisance parameters
we recognise that the input polarisations are consistent with a
circularly polarised wave, so that the signal at the detector is
\begin{align}
\hskip -0.5cm  \tilde{h}_{i}(f) = 
 \biggl[~\frac{1}{2} & F^{(+)}_{i}(\phi,\theta,\psi)(1+\mu^2)\tilde{h}_{+}(f)
   + \nonumber \\
  & F_{i}^{(\times)}(\phi,\theta,\psi)\,\mu\,\tilde{h}_{\times}(f) 
  ~\biggr]~e^{-i\varphi}\,,
\end{align}
where $\phi,\theta,\psi$ are the source right ascension, declination
and polarisation angles, respectively. The cosine of the inclination
angle is given by $\mu$ and we apply an arbitrary constant phase
factor $\varphi$. Explicit definitions of the antenna patterns
$F_{i}^{(+/\times)}$, for the \ac{ET} detector (where $i$ indexes the
three \ac{ET} interferometers) can be found in~\cite{Zhao:2011}.
Independent Gaussian noise $\tilde{n}_{i}(f)$ of spectral amplitude
matching the ET-B noise curve is then added to the frequency domain
waveforms corresponding to each \ac{ET} interferometer.  The simulated
data is then
\begin{equation}
  \tilde{d}_{i}(f_{k}) = \tilde{h}_{i}(f_{k}) + \tilde{n}_{i}(f_{k})\,,
\end{equation}
where $i$ indexes the \ac{ET} interferometers and $k$ the discrete frequency
grid on which the waveform has been evaluated.  This grid is defined by
the length and sampling time of the original numerical waveforms such
that the frequency resolution $\Delta f=(N\Delta t)^{-1} =
31.2118\,\text{Hz}$ and only the $86$ frequencies between $800$ Hz and
$3.5$ kHz are included.

At this point we apply a fifth-order high-pass Butterworth filter to the
data with knee frequency $1$ kHz and a time-domain Tukey window with
alpha parameter $0.25$.  We therefore treat the data in exactly the same
way as done in the template generation procedure and for the same
reasons, namely, to minimise contributions to the signal power from the
last few cycles of the inspiral and merger. It is also important to
perform this filtering \emph{after} the waveform has been redshifted and
noise has been added, since in this way any spectral features resulting
from this pre-processing will not show any artificial cosmological
dependence.

We define the detector noise weighted power as
\begin{equation}
\label{eq:power}
P(f_{k}) \equiv 
4\sum_{i=1}^{3}\frac{|\tilde{d}_{i}(f_{k})|^{2}}{S_{h}^{(i)}(f_{k})}\Delta f.
\end{equation}
At any given frequency, this power is governed by a
non-central $\chi^{2}$ distribution with six degrees of freedom.  The
non-centrality parameter of this distribution is given by the
squared optimal \ac{SNR} in that frequency bin
\begin{equation}
  \rho^{2}(f_{k}) \equiv 
4\sum_{i=1}^{3}\frac{|\tilde{h}_{i}(f_{k})|^{2}}{S_{h}^{(i)}(f_{k})}\Delta
  f\cong\mathcal{S}(f_{k};\bm{\lambda})\,,
\end{equation}
which is approximately equal to our power-spectrum template defined by
Eq.~(\ref{eq:template}).  We note that the freedom in our choice of
amplitude parameters in the template allows us to use an equality rather
than a proportionality in the above relationship.  It then follows that
the likelihood can be written as
\begin{align}
  p(\left\{P\right\}|\bm{\lambda})= 
\prod_{k=1}^{L}\Bigg\{&\frac{P(f_{k})}{2\mathcal{S}(f_{k},\bm{\lambda})}e^{-(\mathcal{S}(f_{k},\bm{\lambda})+P(f_{k}))/2}\nonumber\\
&\times I_{2}(\sqrt{P(f_{k})\mathcal{S}(f_{k},\bm{\lambda})})\Bigg\}\,,
\end{align}
where $I_{2}$ is the second-order modified Bessel function of the first
kind and $L$ is the total number of frequency bins. We note that our
template will now be sensitive to the redshifted frequencies present in
the data and \emph{not} the the intrinsic frequencies.  Therefore, we
define a redshifted parameter vector $\bm{\lambda}_{z}$ containing the
same amplitude parameters as $\bm{\lambda}$, but with the frequencies
$F_{j,z}\equiv F_{j}/(1+z)$, $W_{j,z}\equiv W_{j}/(1+z)$ and $s_{z}
\equiv s/(1+z)$.  Consequently our parameters of interest become the
redshifted characteristic frequencies $f_{j,z}\equiv f_{j}/(1+z)$.

We then apply a nested-sampling algorithm~\cite{Skilling2006} to the data
to provide samples from the posterior distributions of the
$\bm{\lambda}_{z}$ parameter set. We assume uniform priors on all
parameters and use output posterior samples to generate posterior samples
of the redshifted frequencies $f_{1,z}$ and $f_{2,z}$. These samples
represent the posterior \ac{PDF}
$p(f_{j,z}|\{P\})$. We keep the prior parameter ranges of
$\bm{\lambda}_{z}$ constant for all simulations and therefore treat each
simulation identically, independent of system mass and redshift.  The
range of our priors are given in Table~\ref{tab:priors}, where we also define
the additional constraints that the amplitude of the first feature must
be greater than that of the third feature; the slope of the second
feature must be negative; and finally the smoothing length $s_{z}$ must
be less than $10\%$ of the total width of the second feature.

The resultant typical measurement uncertainties in $f_{1}$ and $f_{2}$
for each simulated system as a function of redshift are given
in Table~\ref{tab:freqerror}. These intermediate results show, as
expected, that the accuracy of measurement (reported in brackets) is
$\mathcal{O}(\text{few}\,\%)$ and decreases with increasing distance. It
is also clear that there is a mild trend towards higher percentage
uncertainties in higher-mass systems and that the percentage errors in
frequency are comparable between the two spectral features.

\begin{table}
\begin{tabular}{ccccccccccc}
\hline\hline
& $A_{1}$ & $A_{2a}$ & $A_{2b}$ & $A_{3}$ & $F_{1,z}$ & $F_{2,z}$ & $F_{3,z}$ & $W_{1,z}$
& $W_{2,z}$ &$W_{3,z}$\\
\hline
min & 0 & 0 & 0 & 0 & 1100 & 1700 & 800 & 30 & 150 & 30 \\
max & 100 & 100 & 100 & 100 & 1600 & 2700 & 1500 & 150 & 500 & 500 \\ 
\hline\hline
\end{tabular}
\caption{The range of priors of the model parameters used in the
  calculation of the posterior distributions on $\bm{\lambda}_{z}$.
  Additional prior constraints defined by $A_{1,z}>A_{3,z}$,
  $A_{2a}>A_{2b}$, and $s_{z}<W_{2,z}/5$ are also applied.}\label{tab:priors}
\end{table}
\begin{table}
\begin{tabular}{ccccccc}
\hline\hline
  $z$ & &\multicolumn{5}{c}{Total gravitational mass $M (\Msun)$} \\
  & & 2.6827 & 2.7567 & 2.8325 & 2.9101 & 2.9781\\
  \hline
\multirow{2}{*}{0.01} & & 6.86 (0.6)  & 7.74 (0.6)  & 5.82  (0.4) & 5.03 (0.3) & 6.03 (0.4) \\ 
	& & 4.80 (0.3)  & 7.11 (0.4)  & 8.10 (0.4)  & 10.22 (0.4) & 17.61 (0.7)\\ \\
\multirow{2}{*}{0.02} & & 11.84 (1.0) & 14.59 (1.1) & 13.87 (1.0) & 11.74 (0.8) & 16.26 (1.0) \\ 
	& & 12.87 (0.7) & 18.66 (1.0) & 25.37 (1.2) & 30.87 (1.4) & 52.54 (2.2) \\ \\
\multirow{2}{*}{0.03} & & 19.91 (1.6) & 22.50 (1.8) & 22.91 (1.7) & 22.72 (1.6) & 47.93 (3.1) \\ 
	& & 26.48 (1.5) & 33.60 (1.7) & 51.82 (2.5) & 55.87 (2.5) & 101.56 (4.3) \\ \\
\multirow{2}{*}{0.04} & & 35.09 (2.9) & 38.77 (3.1) & 34.47 (2.6) & 37.21 (2.6) & 67.41 (4.4) \\ 
	& & 44.37 (2.5) & 54.92 (2.9) & 86.99 (4.2) & 84.77 (3.8) & 120.85 (5.1) \\ 
  \hline\hline
\end{tabular}
\caption{The absolute measurement uncertainties in the redshifted
  characteristic frequencies of the dominant \ac{HMNS} spectral
  features. The corresponding percentage uncertainties are given in
  brackets.  We show pairs of results in units of Hz for $f_{1}$ (upper
  rows) and $f_{2}$ (lower rows) for each redshift and for each of the
  five total gravitational masses. Each value represents half of the span
  of the $68\%$ confidence region on the frequency measurement averaged
  over 100 different noise realisations, source and sky
  orientations.}\label{tab:freqerror}
\end{table}
%

\subsection{Measurement of the inspiral phase}\label{sec:inspiral}

We treat the measurement of the inspiral stage of the waveform separately
from the numerical simulations of the \ac{HMNS} stage. At the redshifts
relevant for the \ac{ET} detector, the \ac{SNR} from a \ac{BNS} inspiral
signal is
\begin{equation}
  \langle\rho\rangle \approx 80\,z^{-1}\, \qquad \text{for}\,z<1 \,,
\end{equation}
after averaging over sky position, polarisation and inclination angles.
We ignore the use of tidal information in the inspiral stage for redshift
inference (we discuss this in Sec.~\ref{sec:conclusions}).  Measurement of
just the inspiral phase of the signal allows us to infer the redshifted 
total mass $M_{z},$ but not separately the gravitational mass or redshift.
Given that the \ac{SNR} is going to be very high, the redshifted-mass 
parameters will be very well constrained from the inspiral phase alone. To 
quantify the accuracy of this measurement, we use a Fisher-matrix approach, 
which is a good approximation when the \ac{SNR} is large and identical to that 
used in previous work on \ac{GW} parameter
estimation~\cite{1994PhRvD..49.2658C,2005PhRvD..71h4008A}.  We consider
as our signal model the frequency domain stationary phase approximation
of a non-spinning \ac{BNS} inspiral signal, correct to 3.5 post-Newtonian order. 
It follows that the redshifted chirp mass $\mathcal{M}_{z} \equiv M_{z}\eta^{3/5}$
and symmetric mass ratio $\eta \equiv m_{1}m_{2}/M^{2}$, where $m_{1}$
and $m_{2}$ are the component masses, has fractional measurement
uncertainties of $\Delta\mathcal{M}_{z}/\mathcal{M}_{z}\approx
5\times10^{-6}(z/0.1)$ and $\Delta\eta/\eta\approx 10^{-3}(z/0.1)$ for
$z<1$. This corresponds to a fractional uncertainty in the redshifted total
mass of
\begin{equation}
\label{eq:Mzerr}
  \frac{\Delta{M_{z}}}{M_{z}} \approx \alpha
10^{-3}\left(\frac{z}{0.1}\right)\,,
\end{equation}
which is dominated by the uncertainty in the symmetric mass ratio of 
the system.  In subsequent use of this expression we will not
assume that the system consists of equal component masses.
We also include a constant scale factor $\alpha$ and take a 
conservative approach by setting it equal to 10, therefore 
overestimating the measurement uncertainty of the redshifted 
total mass.

In contrast to the \ac{HMNS} stage, our analysis of the inspiral stage is 
quite simplistic. Given one of our original set of five numerically
evolved binary systems and a choice of redshift, we assume a Gaussian
uncertainty in the measurement of $M_{z}$ based on our Fisher matrix
result. We then draw a random Gaussian variable $M'_{z},$ with mean equal
to the original redshifted total mass and a standard deviation governed
by Eq.~\eqref{eq:Mzerr} with $\alpha=10,$ where a prime stands for
the {\em measured} mass, which includes the measurement uncertainty, as 
opposed to the {\em true} redshifted total mass $M_{z}$.  The 
corresponding likelihood function is then a Gaussian with the same 
standard deviation and mean equal to the randomly drawn $M_{z}$ value.

\subsection{Inference in the $(z,M)$ plane}
\label{sec:inference}

Starting with the measurement of the inspiral phase described in the 
previous section, we can write the joint posterior \ac{PDF} of the total
gravitational mass and redshift as
\begin{equation}\label{eq:inspiraljoint}
  p(z,M|M'_{z}) \propto 
\exp\left(-\frac{(M'_{z}-M(1+z))}{2(\Delta M_{z})^{2}}\right)\,.
\end{equation}
In the $(z,M)$ plane this defines a narrow ``stripe'' of probability
spanning a region that extends from high masses at low redshift, to low
masses at high redshift (see the blue curve in Fig.~\ref{fig:sketch}).
We have assumed flat prior distributions for $M$ and $z$.

For the \ac{HMNS} measurement, the following procedure is applied to each
spectral feature indexed using $j$.  We note that any given
redshift and total gravitational mass of a \ac{BNS} system defines a
specific point in the $(z,M)$ plane. From Eq.~\eqref{eq:polyfit} we can
determine the intrinsic characteristic frequency of the spectral
feature in question at this value of $M$.  We can also 
calculate the corresponding redshifted characteristic frequency using the
relation $f_{j,z} = f_{j}/(1+z)$. The probability density associated with
obtaining any particular value of $f_{j,z},$ hence that particular
$(z,M)$ pair, is given by the marginalised posterior \ac{PDF},
$p(f_{1,z}|\{P\})$, obtained from our analysis of the \ac{HMNS} data described
in Sec.~\ref{sec:barmode}. We can, therefore, write
\begin{equation}
\label{eq:bmijoint}
  p_{j}(z,M|\{P\}) \propto 
p\left(f_{j,z}=\frac{f_{j}(M)}{(1+z)}\Bigg\vert\{P\}\right)\,.
\end{equation}
This function will describe an arc of probability in the $(z,M)$ plane
that sweeps almost orthogonally to that obtained from the measurement of 
the inspiral phase. An increase in the observed redshifted frequency can
be obtained via either an increase in the total mass of the binary or a 
decrease in redshift of the source.  Consequently, a high-redshift 
high-mass system will generally have similar redshifted characteristic 
frequencies to a low-redshift low-mass system.

\begin{figure*}
\centering
\includegraphics[width=\textwidth]{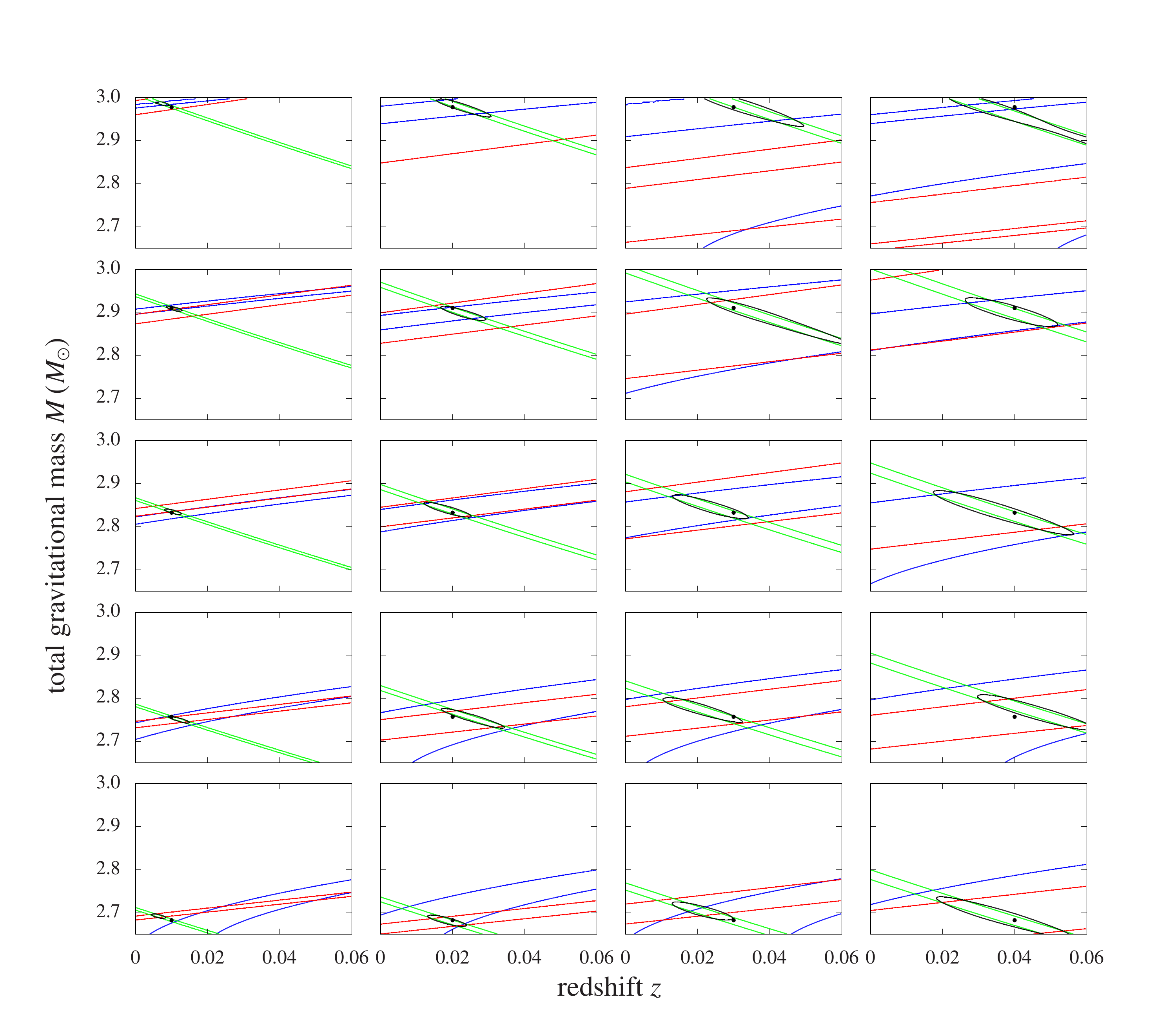}
\caption{The joint posterior distributions on the redshift and total
  gravitational mass of the \ac{BNS} system for single representative
  realisations of noise and system parameters. Each row of plots
  represents a simulated signal of one of the five system masses (see
  Table~\ref{tab:ID}) ranging from low (bottom row) to high (top row)
  mass.  Columns represent different simulated redshifts ranging from
  $0.01$ (left) to $0.04$ (right) in steps of $0.01$.  The green, blue
  and red contours represent the posterior contributions from the
  inspiral measurement, the first \ac{HMNS} spectral feature and the
  second \ac{HMNS} spectral feature respectively. The black contours
  represent the final posterior distribution combining all measurements
  and the black dots indicate the true simulated redshift and total mass
  values. In all cases the contours enclose $68\%$ of the probability.}
\label{fig:zM_multimap}
\end{figure*}

Combining the information from both spectral features and the inspiral
phase of the signal and assuming statistical independence, the final joint
posterior distribution of $z$ and $M$ is simply the product of all three
distributions
\begin{equation}\label{eq:joint}
  p(z,M|M'_{z},\{P\})\propto p(z,M|M'_{z})\prod_{j=1}^{2}p_{j}(z,M|\{P\})\,.
\end{equation}
We show examples of joint posterior probabilities of $z$ and $M$ for all five
systems studied, and for multiple redshift values, in
Fig.~\ref{fig:zM_multimap}. These examples mirror the original conceptual
sketch in Fig.~\ref{fig:sketch}, where we can see how, in practice,
the mass-redshift degeneracy is broken through the use of the spectral
properties of the \ac{HMNS} \ac{GW} signal.  For sources at redshifts $z=0.01$--$0.04$ 
the uncertainty in the measurement of the redshift is $\Delta z \sim 
10\%$--$20\%,$ over the full range of simulated system masses. In
addition, the gravitational mass can be measured with fractional
errors of $<1\%$ in all cases.

In Tables~\ref{tab:zerror} and~\ref{tab:Merror} we give representative
values for the fractional uncertainty on the measured redshift and total
gravitational mass, as a function of their true simulated values.  These
uncertainty estimates are obtained by marginalising the joint
distribution $p(z,M|M'_{z},\{P\})$ over the total mass to obtain
$p(z|M_{z},\{P\})$, and over the the redshift to obtain
$p(M|M_{z},\{P\})$. From these marginalised distributions we compute our
representative uncertainties as half of the minimum interval to contain
$68\%$ of the total probability (analogous to the 1-sigma uncertainty for
a Gaussian distribution).

\begin{table*}
\begin{tabular}{ccccccccccccccccccccc}
\hline\hline
  $z$ & &\multicolumn{19}{c}{Total gravitational mass $M$($\Msun$)} \\
  & & \multicolumn{3}{c}{2.6827} & & \multicolumn{3}{c}{2.7567} &
  &\multicolumn{3}{c}{2.8325} &  &\multicolumn{3}{c}{2.9101} & & \multicolumn{3}{c}{2.9781}\\
  \hline
0.01 & & 39.1 & 7.8 & \bf{7.5} & & 24.6 & 10.6 & \bf{9.7} & & 13.5 & 11.2 & \bf{8.5} & & 9.6 & 13.8 & \bf{7.9} & & 9.4 & 20.7 & \bf{8.7} \\ 
0.02 & & 40.4 & 10.0 & \bf{9.2} & & 25.7 & 13.9 & \bf{11.6} & & 17.5 & 16.5 & \bf{11.9} & & 11.6 & 20.6 & \bf{10.1} & & 13.4 & 32.8 & \bf{12.0} \\ 
0.03 & & 37.5 & 13.6 & \bf{11.6} & & 28.1 & 17.1 & \bf{14.2} & & 18.8 & 21.3 & \bf{14.7} & & 15.0 & 25.0 & \bf{13.0} & & 29.1 & 36.4 & \bf{19.6} \\ 
0.04 & & 31.8 & 15.3 & \bf{14.9} & & 26.8 & 19.1 & \bf{15.8} & & 21.2 & 23.5 & \bf{15.7} & & 18.6 & 26.4 & \bf{15.1} & & 23.7 & 30.5 & \bf{19.4} \\
 \hline\hline
\end{tabular}
\caption{The percentage measurement uncertainties on the redshift of a
  \ac{BNS} source. We show results for each of our five different mass
  systems and for each of four different redshifts. We give three
  fractional redshift uncertainties for each combination. The first and
  second correspond to results using the inspiral measurement plus the
  first and second spectral features respectively. The third result (in
  boldface) is from a combination of the inspiral measurement and both
  spectral features. We have performed analyses of 100 different noise
  realisations, source and sky orientations for each redshift and mass
  combination.  For each realisation we compute a quantity equal to half
  of the span of the $68\%$ confidence region on the redshift
  measurement.  The quoted value is the median of this quantity over the
  100 realisations.\label{tab:zerror}}
\end{table*}
\begin{table*}
\begin{tabular}{ccccccccccccccccccccc}
\hline\hline
  $z$ & &\multicolumn{19}{c}{Total gravitational mass $M$($\Msun$)} \\
  & & \multicolumn{3}{c}{2.6827} & & \multicolumn{3}{c}{2.7567} &
  &\multicolumn{3}{c}{2.8325} &  &\multicolumn{3}{c}{2.9101} & & \multicolumn{3}{c}{2.9781}\\
  \hline
0.01 & & 0.38 & 0.07 & \bf{0.06} & & 0.24 & 0.10 & \bf{0.09} & & 0.13 & 0.10 & \bf{0.07} & & 0.09 & 0.13 & \bf{0.07} & & 0.08 & 0.20 & \bf{0.07} \\ 
0.02 & & 0.78 & 0.18 & \bf{0.17} & & 0.50 & 0.26 & \bf{0.22} & & 0.33 & 0.32 & \bf{0.22} & & 0.21 & 0.39 & \bf{0.18} & & 0.24 & 0.63 & \bf{0.22} \\ 
0.03 & & 1.06 & 0.38 & \bf{0.33} & & 0.80 & 0.48 & \bf{0.40} & & 0.53 & 0.62 & \bf{0.41} & & 0.42 & 0.73 & \bf{0.36} & & 0.79 & 1.04 & \bf{0.55} \\ 
0.04 & & 1.24 & 0.61 & \bf{0.58} & & 1.03 & 0.72 & \bf{0.60} & & 0.80 & 0.93 & \bf{0.60} & & 0.68 & 0.98 & \bf{0.55} & & 0.87 & 1.11 & \bf{0.71} \\ 
 \hline\hline
\end{tabular}
\caption{The percentage measurement uncertainties on the total
  gravitational mass of a \ac{BNS} source. We show results for each of our
  five different mass systems and for each of four different redshifts.
  The details are identical to those given in
  Table~\ref{tab:zerror}.\label{tab:Merror}}
\end{table*}
%

\section{Conclusions}\label{sec:conclusions}
%

A well-known problem of the detection of \acp{GW} from compact-object binaries
at cosmological distances is the so-called mass-redshift degeneracy,
namely, that \ac{GW} measurements allow the determination of the redshifted
mass $M_{z}=M(1+z)$, but not the gravitational mass $M$ or the redshift
$z$ separately. \ac{GW} observations allow the measurement of the luminosity
distance but this degeneracy restricts the cosmological application of GW
observations, since it is the relation between the source's luminosity
distance and its redshift that allows us to probe cosmological models.
Until recently it was thought that coincident \ac{EM} and \ac{GW} 
observations would be required to break the mass-redshift degeneracy,
as \ac{EM} observations would allow the host galaxy to be identified
and hence the extraction of the redshift. In this paper we have
described a novel approach to this problem that does not require an
electromagnetic counterpart and exploits instead the information encoded
in the \ac{HMNS} stage of a \ac{BNS} waveform to break the mass-redshift
degeneracy.

We have described how, with the use of five numerically generated \ac{BNS}
waveforms of very slightly differing mass, we have been able to construct
frequency-domain power-spectrum reference templates.  The templates were
designed to capture the evolution of two primary spectral features in
the \ac{HMNS} stage of the waveforms, as a function of the total gravitational
mass.  The characteristic frequencies of these spectral features were
then fitted to polynomial functions of mass providing us with an ad-hoc
approximation to the characteristic frequencies for any mass. A 
Bayesian inference method was used to test the ability of the \ac{ET}
to measure the characteristic frequencies in the \ac{HMNS} stage 
of the signal. These frequency measurements were coupled with our 
precomputed, empirical, frequency-mass relation and a measurement of 
the redshifted mass from the inspiral phase of the signal, allowing us 
to determine both the redshift and gravitational mass separately.

We have shown that in an analysis based on the signal's
power spectrum, and ignoring all phase information within the \ac{HMNS}
stage, the measurement uncertainties in the redshift of sources at
$z=0.01$-$0.04$ is $\sim 10\%$-$20\%,$ over the full range of
simulated system masses.  In addition we find that the gravitational
mass can be measured with fractional accuracies of $<1\%$ in all cases. 

We have specifically ignored the tidal effects in the late insprial phase
that have been previously shown to be useful in redshift measurements.
This choice was made to simplify our analysis and clearly identify the
potential of this new technique.  It is encouraging to find that the 
two approaches both have comparable redshift sensitivities of $\sim
\mathcal{O}(10\%)$, implying that a combination of their results will
improve the overall redshift estimate.

Under the hypothesis that there is a single universal \ac{NS} \ac{EOS}, it is
highly likely that by the time of \ac{ET} the \ac{NS} \ac{EOS} will be
tightly constrained via various observations including direct GW
detections from the advanced \ac{GW} detectors.  We have limited our 
study to a single \ac{EOS}, but based on previous
studies~\cite{Bauswein2011,Bauswein2012} would expect that the general
result holds for all realistic possibilities.  Of primary interest here
is the general concept that there exists an additional ``matter-effect''
found in the \ac{HMNS} stage of the waveform that can provide frequency
markers from which redshift information can be obtained.

We should stress that this analysis is one of the first attempts to
perform parameter estimation on the \ac{HMNS} stage of \ac{BNS} signals.  Whilst
the numerically generated waveforms and Bayesian parameter estimation
techniques used here represent the state-of-the-art, our analytic signal
model approximation and mass-frequency fitting is necessarily simplistic. 
We expect that prior to the era of third generation \ac{GW} detectors, the 
understanding of \ac{BNS} mergers through numerical relativity and direct 
GW detections will enable us to significantly enhance our ad-hoc models.  
In the future, a significant improvement in
the accuracies of redshift measurements using the \ac{HMNS} stage could become
possible if a realistic model of the phase evolution were constructed. In
such a scenario, an analysis of the type presented here may be applicable
to signals found in the advanced \ac{GW} detectors.

\section*{Acknowledgements}

We would like to acknowledge the useful discussions with our
colleagues from the LSC-Virgo Collaboration and CM, SG and BSS
especially thank C. Ott, J. Read and J. Veitch. CM and BSS
were funded by the Science and Technology Facilities Council (STFC)
Grant No. ST/J000345/1 and SG is partially supported by NSF grants
PHY-1151197 and 1068881.  This work was supported in part by the DFG
grant SFB/Transregio~7 and by ``CompStar'', a Research Networking
Programme of the European Science Foundation. The simulations were
performed on SuperMUC at LRZ-Munich, on Datura at AEI-Potsdam and on
LOEWE at CSC-Frankfurt.

\bibliographystyle{apsrev4-1-noeprint.bst}
\bibliography{Bibliography}
\end{document}